# Temperature-dependent structure of an intermetallic ErPd$_2$Si$_2$ single crystal: A combined synchrotron and in-house X-ray diffraction study

Kaitong Sun,[1, 2, 3, *] Yinghao Zhu,[1, 3, *] Si Wu,[1, 3, *] Junchao Xia,[1] Pengfei Zhou,[1] Qian Zhao,[1] Chongde Cao,[4, 5, †] Hai-Feng Li[1, ‡]

[1]*Joint Key Laboratory of the Ministry of Education, Institute of Applied Physics and Materials Engineering, University of Macau, Avenida da Universidade, Taipa, Macao SAR 999078, China.*
[2]*State Key Laboratory of High Performance Ceramics and Superfine Microstructure, Shanghai Institute of Ceramics, Chinese Academy of Sciences, Shanghai 200050, China.*
[3]*Guangdong–Hong Kong–Macao Joint Laboratory for Neutron Scattering Science and Technology, No. 1. Zhongziyuan Road, Dalang, DongGuan 523803, China.*
[4]*Research and Development Institute of Northwestern Polytechnical University in Shenzhen, Shenzhen 518057, China.*
[5]*School of Physical Science and Technology, Northwestern Polytechnical University, Xian 710072, China.*
(Date: March 28, 2022)

We have grown intermetallic ErPd$_2$Si$_2$ single crystals employing laser-diodes with the floating-zone method. The temperature-dependent crystallography was determined using synchrotron and in-house X-ray powder diffraction measurements from 20 to 500 K. The diffraction patterns fit well with the tetragonal *I4/mmm* space group (No. 139) with two chemical formulas within one unit cell. Our synchrotron X-ray powder diffraction study shows that the refined lattice constants are $a$ = 4.10320(2) Å, $c$ = 9.88393(5) Å at 298 K and $a$ = 4.11737(2) Å, $c$ = 9.88143(5) Å at 500 K, resulting in the unit-cell volume $V$ = 166.408(1) Å$^3$ (298 K) and 167.517(2) Å$^3$ (500 K). In the whole studied temperature range, we did not find any structural phase transition. Upon cooling, the lattice constants $a$ and $c$ are shortened and elongated, respectively.

## I. INTRODUCTION

The ThCr$_2$Si$_2$-type structure is one of the most abundant structural prototypes of ternary intermetallics (Bażela et al., 1997; Cao et al., 2008; Frontzek 2009; Shatruk 2019), which shows a variety of interesting physical properties such as pressure-induced superconductivity in CePd$_2$Si$_2$ (Stewart 2001), anomalous valence fluctuations in EuPd$_2$Si$_2$ (Stewart 2001), and heavy fermion behavior (Shatruk 2019). Among them, the Erbium palladium silicide (ErPd$_2$Si$_2$) displays an anisotropic magnetic behavior along the three crystallographic axes at low temperatures (Sampathkumaran et al., 2008). The ErPd$_2$Si$_2$ compound shows ferromagnetic ordering along the *c* axis and antiferromagnetic ordering within the *ab* plane, according to the measurements of paramagnetic Curie-Weiss behavior derived from the high-temperature linear regime of the inverse magnetic susceptibility $\chi^{-1}$ (Sampathkumaran et al., 2008). Furthermore, the magnetic susceptibility $\chi$ parallel to the *ab* plane shows a sharp peak at ~ 3.8 K and a board peak in the temperature range of 8–20 K. These were attributed to spin fluctuations with a finite antiferromagnetic component (Sampathkumaran et al., 2008). Moreover, a neutron powder diffraction study on polycrystalline ErPd$_2$Si$_2$ reveals a sine modulated magnetic structure (Bażela et al., 1991). Later, polarized and unpolarized neutron diffraction studies reported two distinct antiferromagnetic modulations with respective propagation vectors at $Q_\pm = (H \pm 0.557(1), 0, L \pm 0.150(1))$ and $Q_C = (H \pm 0.564(1), 0, L)$ for a ErPd$_2$Si$_2$ single crystal (Li et al., 2015). The $Q_\pm$ modulation was attributed to localized 4*f* moments, whereas the $Q_C$ was related to itinerant moments from conduction bands (Li et al., 2015).

Intermetallic *RE*Pd$_2$Si$_2$ (*RE* = rare earth) silicides (Sampathkumaran et al., 1984; Mazilu et al., 2008; Frontzek 2009; Stewart 2001; Prokofiev 2018; Xu et al., 2011) have the same tetragonal *I4/mmm* structure as the high-temperature structural phase of the family of 122-iron-pnictides (Sasmal et al., 2008; Li et al., 2009; Li et al., 2010). For example, there exist simultaneous antiferromagnetic and structural phase transitions for SrFe$_2$As$_2$ at ~ 201.5 K upon cooling, from the high-temperature tetragonal phase (*I4/mmm*) to the low-temperature orthorhombic structure (*Fmmm*). With hole or electron doping in the parent SrFe$_2$As$_2$, superconductivity appears with $T_C$ up to 55 K after both the magnetic and structural phase transitions were suppressed (Rotter et al., 2008; Sefat et al., 2008; Torikachvili Sefat et al., 2008).

Most of the previous studies on ErPd$_2$Si$_2$ were focused on the low-temperature regime (Yakinthos and Gamari-Seale 1982; Bażela et al., 1991; Tomala et al., 1994; Bażela et al., 1997; Szytuła et al., 2001; Cao et al., 2014; Uchima et al., 2018), and there exist few temperature-dependent studies, especially from the structural point of view. Therefore, we grew the intermetallic ErPd$_2$Si$_2$ single crystals and performed a temperature-dependent structural study up to 500 K. We carried out both in-house X-ray powder diffraction (XRPD) and synchrotron X-ray powder diffraction (SXRPD) measurements on a pulverized ErPd$_2$Si$_2$ single crystal to determine its temperature-dependent crystallographic information and check potential structural phase transition.

## II. EXPERIMENTAL
### A. Single crystal growth

First, we prepared polycrystalline ErPd$_2$Si$_2$ samples with constituent metals of Er (99.98% purity), Pd (99.95% purity), and Si (99.99% purity) in a well-equipped arc-melting furnace (WK-11, Physcience Opto-electronics Co., Ltd.) under argon atmosphere (99.999% purity) at the University of Macau, Macao, China. Initially, additional 3–5% mole Pd metal was added according to the stoichiometric ratio to supplement the loss due to volatilization. Having melted the mixture for three times, the ingot was ground manually, and the resulting powder was filled into plastic balloons for preparations of seed and feed rods. The balloon was shaped with a hydrostatic pressure of ~ 70 MPa. The prepared rods were sealed into cylindrical glass tubes with protecting argon gas and sintered at 950 °C for 36 h. After sintering, the samples are pure $ErPd_2Si_2$ phase without additional metals or impurity oxides. We grew the single crystals of $ErPd_2Si_2$ compound with the sintered rods treated additionally (Li et al., 2021) by the floating-zone (FZ) technique using a well-equipped laser-diode FZ furnace (Model: LD-FZ-5-200W-VPO-PC-UM) at the University of Macau, Macao, China (Wu et al., 2020). The growth conditions are similar to those reported previously (Cao et al., 2014). It is stressed that the grinding and the shaping processes were carried out in a glove box. Scanning electron microscopy with energy dispersive X-ray analysis of the grown $ErPd_2Si_2$ single crystal reveals a chemical stoichiometry of $Er_{1.00(5)}Pd_{2.10(7)}Si_{2.10(10)}$ (Figure 1), indicating ~ 5% vacancies on the Er site.

**B. High-temperature synchrotron X-ray powder diffraction**

We gently ground a $ErPd_2Si_2$ single crystal into powdered sample using a Vibratory Micro Mill (FRITSCH PULVERISETTE 0) with a vertical vibrating amplitude of 0.5 mm for 1.0 h for the SXRPD study to determine the high-temperature crystallographic information of $ErPd_2Si_2$. The grain size of the powdered sample is 4.07 ± 0.97 $\mu m$ on average. The SXRPD measurements were performed on the beamline I11 at Diamond Light Source, Didcot, UK (Thompson et al., 2009; Li et al., 2014; Tang et al., 2015). The beamline consists of an array of permanent magnets, three sets of slits, a double-crystal-monochromator comprising two liquid nitrogen cooled Si (111) crystals, a pair of double bounce harmonic rejection mirrors, and an intensity monitor comprising a thin Kapton scattering foil and scintillation counter. The beamline comprises a transmission geometry X-ray instrument with a wide range of position sensitive detectors. We used a triple-axis/two-circle diffractometer with high precision rotary stages. Synchrotron X-rays with a wavelength of $\lambda$ = 0.827032 Å were chosen as the radiation source. High-resolution SXRPD patterns were collected over a diffraction $2\theta$ angle range of 9–66° at 298 and 500 K. The $2\theta$ step interval is 0.001°, and the counting time is 1800 s (for 298 and 500 K) and 3600 s (for 500 K). The $ErPd_2Si_2$ powder was loaded onto the external surface of a 0.3-mm diameter borosilicate glass capillary tube by applying a thin layer of hand cream to it. The capillary sample holder was mounted directly on a magnetic spinner and in the center of the $\theta$ circle faceplate. The magnetic spinner keeps rotating around the vertical axis of the sample holder plate during the SXRPD measurements. The rotation technique is effective to obtain quality diffraction profiles and to minimize absorption and issue of preferred orientation during data collection. The sample temperature was controlled using the beamline hot air blower and detected by a thermocouple near the sample. The symmetric $\theta$-$2\theta$ scan technique was used for data collection (Thompson et al., 2009; Tang et al., 2015).

**C. Low-temperature in-house X-ray powder diffraction**

We used the same powdered ErPd$_2$Si$_2$ sample for an in-house X-ray powder diffraction study. The measurements were carried out on an in-house X-ray diffractometer (Rigaku, SmartLab 9 kW) employing cooper $K_{\alpha 1}$ = 1.54056 Å as the radiation and a 2D multidimensional semiconductor detector. XRPD patterns were collected at a voltage of 45 kV and a current of 200 mA. The $2\theta$ range is from 30 to 78° with a step size of 0.005°. The measurements were performed at 20, 100, and 200 K with a dwell time of 1 h at each temperature.

**D. Rietveld refinements**

The computer program FULLPROF SUITE (Rodríguez-Carvajal 1993) was used to analyze all powder diffraction data by estimating the exact $2\theta$ position for each peak. The initial crystal structure model was referred to the room-temperature XRPD study (Li et al., 2015). The peak profile shape was modeled with a Pseudo-Voigt function. We refined the background using linear interpolation between automatically identified background points. We considered the parameters relevant to improving refinement results, including scale factor, zero shift, peak shape parameters, asymmetry, preferred orientation, lattice constants, atomic positions, as well as isotropic thermal parameter $B$. The refining procedure is as follow: (i) First, we refined the scale factor, zero shift, and lattice constants. (ii) Second, we refined the peak shape parameters and background points. (iii) In the following, we refined the atomic positions, thermal parameters, asymmetry, and preferred orientation step by step. Furthermore, we tentatively refined the strain effect on the collected data with a general strain broadening model (quartic form).

**III. RESULTS AND DISCUSSION**

**A. High-temperature structure from SXRPD study**

Compared to in-house X-ray powder diffraction and neutron powder diffraction based on large facilities, SXRPD holds the highest resolution for studying structural phase transitions by detecting possible Bragg-peak splitting and observing additional Bragg peaks at higher scattering angles. Figure 2 displays the SXRPD patterns of ErPd$_2$Si$_2$ measured at 298 K for 0.5 h (Figure 2(a)) and 500 K for 0.5 h (Figure 2(b)) and for 1.0 h (Figure 2(c)). All observed patterns were well refined with the tetragonal $I4/mmm$ model. We did not find any forbidden Bragg peak (Insets of Figure 2). Table I lists values of refinement reliability parameters, $R_p$, $R_{wp}$, and $R_{exp}$, as well as the goodness of fit $\chi^2$. These values are acceptable within the present experimental accuracy, for example, the $\chi^2$ = 1.74 at 298 K and 2.12 at 500 K, which validates our FULLPROF refinements. By carefully checking the patterns at 298 K (Figure 2(a)) and 500 K (Figures 2(b) and 2(c)), we confirm that there is no structural phase transition as temperature increases from 298 to 500 K.

The extracted unit cell of the refined structural model was exhibited in Figure 3, where Er, Pd, and Si ions were marked. It is noted that the Pd and Si layers are staggered with a separation of Er layer. There is only one degree of freedom for the change in atomic positions of Er, Pd, and Si ions, i.e., the $z$ position of Si ions. Therefore, the electric and magnetic properties of ErPd$_2$Si$_2$ compound, e.g., anisotropic magnetoresistance, were determined critically by the Er ions and their coupling between interlayers, as well as the position of Si ions.

The splitting degree of Bragg peaks will become larger and larger as diffraction angle increases. Therefore, the diffraction data at higher $2\theta$ angles is better for monitoring possible structural phase transitions. Figure 4 shows SXRPD patterns as well as the corresponding FULLPROF refinements in the $2\theta$ range of 50.3–54.8°. The Bragg peaks are indexed as (4 1 3), (3 3 2), (4 0 4), (2 1 9), (2 2 8), (1 1 10), and (4 2 0) as marked in Figure 4(b). The refinements for 298 and 500 K data coincide well with the splitting of (4 1 3), (3 3 2) and (4 0 4) Bragg peaks.

We collected the SXRPD data at 500 K for 0.5 h (Figure 2(b)) and for 1.0 h (Figure 2(c)) to check the effect of counting time on Rietveld refinements. As listed in Table I, the reliability values of $R_p$, $R_{wp}$, and $R_{exp}$ for the data collected at 500 K with counting time of 0.5 h are higher than those of the data with counting time of 1.0 h, indicating that a higher counting time increases the ratio of intensity/background. Whereas the goodness of fit ($\chi^2$ = 1.64) of 0.5 h data is smaller than that (2.12) of the 1.0 h data, which we attribute to the enhancement of diffracted intensity. Finally, the data with a longer counting time also did not show peak splitting and additional Bragg peaks.

We listed the refined unit cell parameters of all patterns in Table II. The lattice constants of ErPd$_2$Si$_2$ were extracted as $a$ = 4.10320(2) Å at 298 K and 4.11737(2) Å at 500 K, displaying an increase with temperature. The lattice constant $c$ equals to 9.88393(5) Å at 298 K and 9.88143(5) Å at 500 K, showing a decrease upon warming. The unit-cell volume $V$ = 166.408(1) Å$^3$ at 298 K and 167.517(2) Å$^3$ at 500 K, expanded by ~ 0.67% with temperature. This is accompanied by a decrease in density from 8.706 g•cm$^{-3}$ (298 K) to 8.649 g•cm$^{-3}$ (500 K). The refined results show a lattice shrinkage along the $c$ axis and an extension along the $a$ axis upon warming. The tetragonal crystal system does not change in the whole studied temperature range. The detailed synchrotron X-ray powder diffraction data at 298 K was indexed in Table III.

In the ThCr$_2$Si$_2$-type structure (Figure 3), Er ions are located at the Wyckoff site 2a (0, 0, 0), and Pd ions are fixed at 4d (0, 0.5, 0.25). The Si ions stay at 4e (0, 0, $z$) with only one degree of freedom along the coordination $z$ axis. The $z$ coordinates of Si ions were calculated as 0.38064(14) at 298 K and 0.37984(13) at 500 K.

**B. Low-temperature structure from in-house XRPD study**

Figure 5 shows the XRPD diffraction patterns collected at 20 K (Figure 5(a)), 100 K (Figure 5(b)), and 200 K (Figure 5(c)), as well as the corresponding structural refinements. We meticulously examined the temperature evolution of the shape and position of Bragg peaks. Overall, the collected data can be effectively indexed with the *I*4/*mmm* space group. The extracted crystallographic information, such as unit cell parameters and atomic positions as well as the goodness of fit, is listed in Table IV. The lattice constants *a* and *c* increase by ~ 0.17% and decrease by ~ 0.11%, respectively, as the temperature rises from 20 to 200 K. Temperature variances in lattice constants cause an expansion of the unit-cell volume and a decrease of the calculated density by ~ 0.23% with temperature.

**C. Strain effect**

Although the values of goodness of fit are low enough for a good FULLRPOF refinement, for example, $\chi^2$ = 1.74 for the refinement of SXRPD data collected at 298 K (Table I), and $\chi^2$ = 1.43 for the refinement of XRPD data collected at 100 K (Table IV), we still tried to refine the strain parameters with an introduction of a general strain broadening model (quartic form) for the collected data. As listed in Table I, the values of reliability parameters and goodness of fit get a little smaller with refinements including the strain parameters, indicating small strain exists in the pulverized powder sample.

If the quality of powder diffraction data was not good enough, sometimes, one may get negative isotropic thermal parameter, which is physical nonsense. It is well known that the isotropic thermal parameter *B* is strongly coupled with background and the relevant occupation factor. Refining powder diffraction data from different measurements, one may extract different values of thermal parameters. It is pointed out that we present our SXRPD and XRPD data as well as the corresponding analyses independently, and no normalization was performed for the unit cell parameters.

**IV. CONCLUSION**

We have synthesized ErPd$_2$Si$_2$ single crystals by a laser-diode FZ furnace. We performed SXRPD and in-house XRPD studies on a pulverized ErPd$_2$Si$_2$ single crystal from 20 to 500 K. The FULLPROF refinements demonstrate that there is no structural phase transition in the whole studied temperature range, and the space group keeps in the tetragonal *I*4/*mmm* system. The refined unit cell parameters display an elongation in the basal *ab* plane and a shrinkage along the *c* axis, resulting in a larger and larger unit-cell volume upon warming. The research provides temperature-dependent crystallographic information of single-crystal ErPd$_2$Si$_2$ compound, which would serve as an important basis for further experimental and theoretical studies.

**V. DEPOSITED DATA**

CIF and/or RAW data files were deposited with ICDD. You may request this data from ICDD at info@icdd.com.

**ACKNOWLEDGMENTS**

The work at University of Macau was supported by the opening project of State Key Laboratory of High Performance Ceramics and Superfine Microstructure (Grant No. SKL201907SIC), Science and Technology Development Fund, Macao SAR (File Nos.

0051/2019/AFJ and 0090/2021/A2), Guangdong Basic and Applied Basic Research Foundation (Guangdong–Dongguan Joint Fund No. 2020B1515120025), University of Macau (MYRG2020-00278-IAPME and EF030/IAPME-LHF/2021/GDSTIC), and Guangdong–Hong Kong–Macao Joint Laboratory for Neutron Scattering Science and Technology (Grant No. 2019B121205003).

The work at Northwestern Polytechnical University was supported by Shenzhen Fundamental Research Program (202103243003450), the National Natural Science Foundation of China (51971180), Shaanxi Provincial Key R&D Program (2021KWZ-13), and Guangdong Provincial Key R&D Program (2019B090905009).

**CONFLICTS OF INTEREST**

The authors have no conflicts of interest to declare.

\* These authors contributed equally.
† caocd@nwpu.edu.cn
‡ haifengli@um.edu.mo

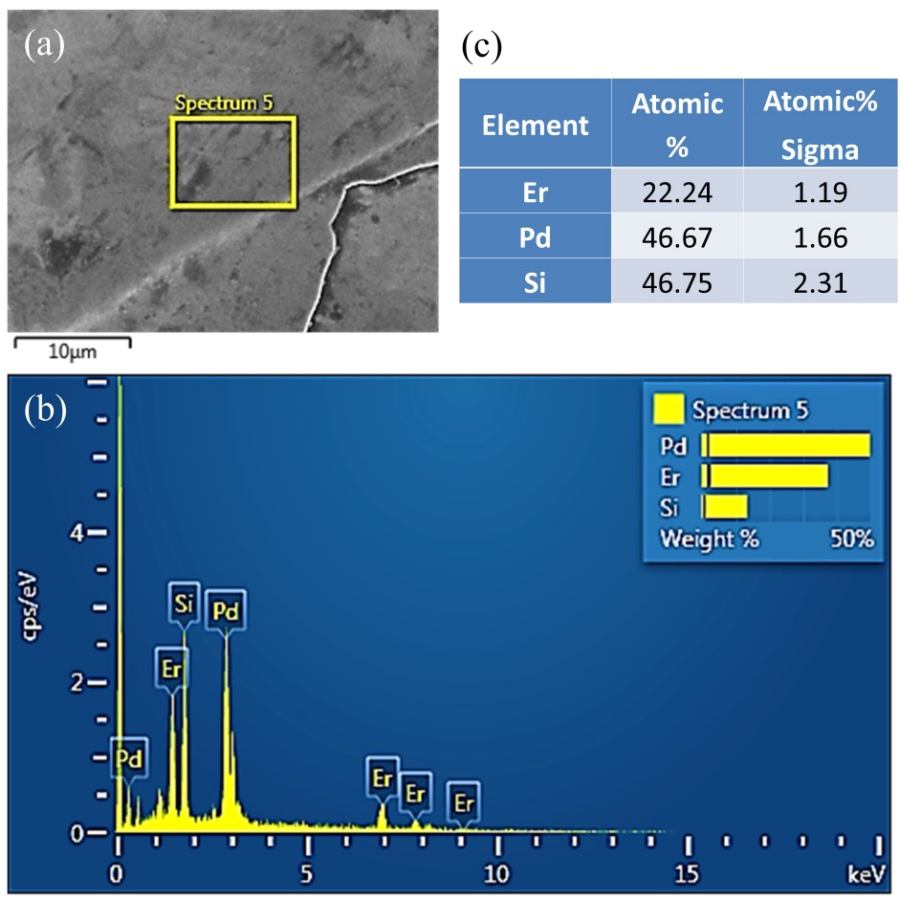

FIG. 1. (a) Scanning electron microscopy image of the ErPd$_2$Si$_2$ single crystal with a scale bar of 10 $\mu$m. (b) Energy dispersive X-ray analysis of the ErPd$_2$Si$_2$ single crystal. (c) Atomic ratio of Er, Pd, and Si elements in the ErPd$_2$Si$_2$ single crystal.

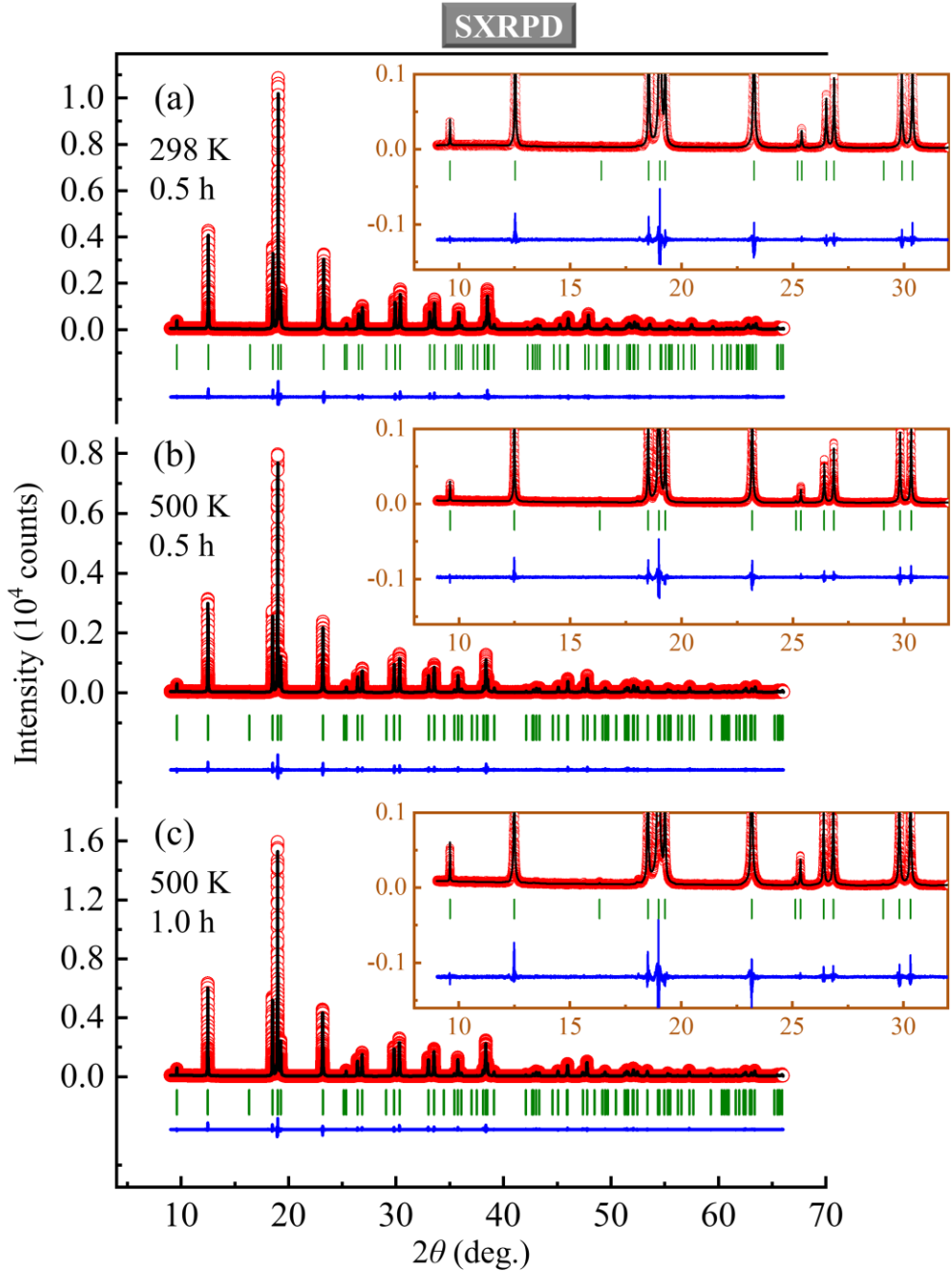

FIG. 2. Observed (circles) and calculated (solid lines) synchrotron X-Ray powder diffraction patterns of a pulverized ErPd$_2$Si$_2$ single crystal, collected at (a) 298 K, 0.5 h; (b) 500 K, 0.5 h; and (c) 500 K, 1.0 h. Vertical bars mark the positions of Bragg reflections. The bottom curves represent the difference between observed and calculated patterns. The diffraction angle $2\theta$ is in the range of 9–66°. Insets of (a), (b) and (c) display their respective enlarged patterns in the $2\theta$ range of 8–32°.

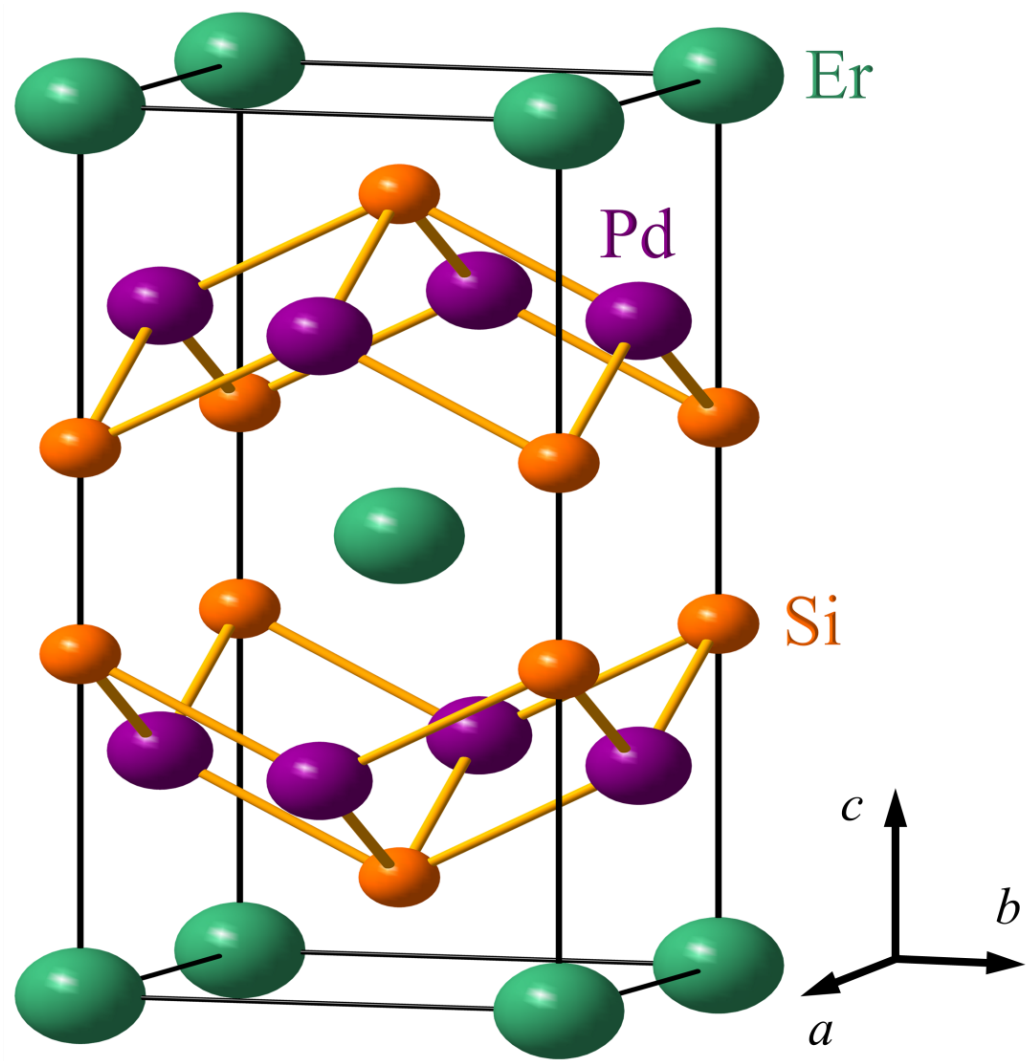

FIG. 3. Refined crystal structure of the single-crystal ErPd$_2$Si$_2$ in one unit cell (solid lines) with space group of $I4/mmm$ (No. 139). The Er, Pd, and Si ions are labeled.

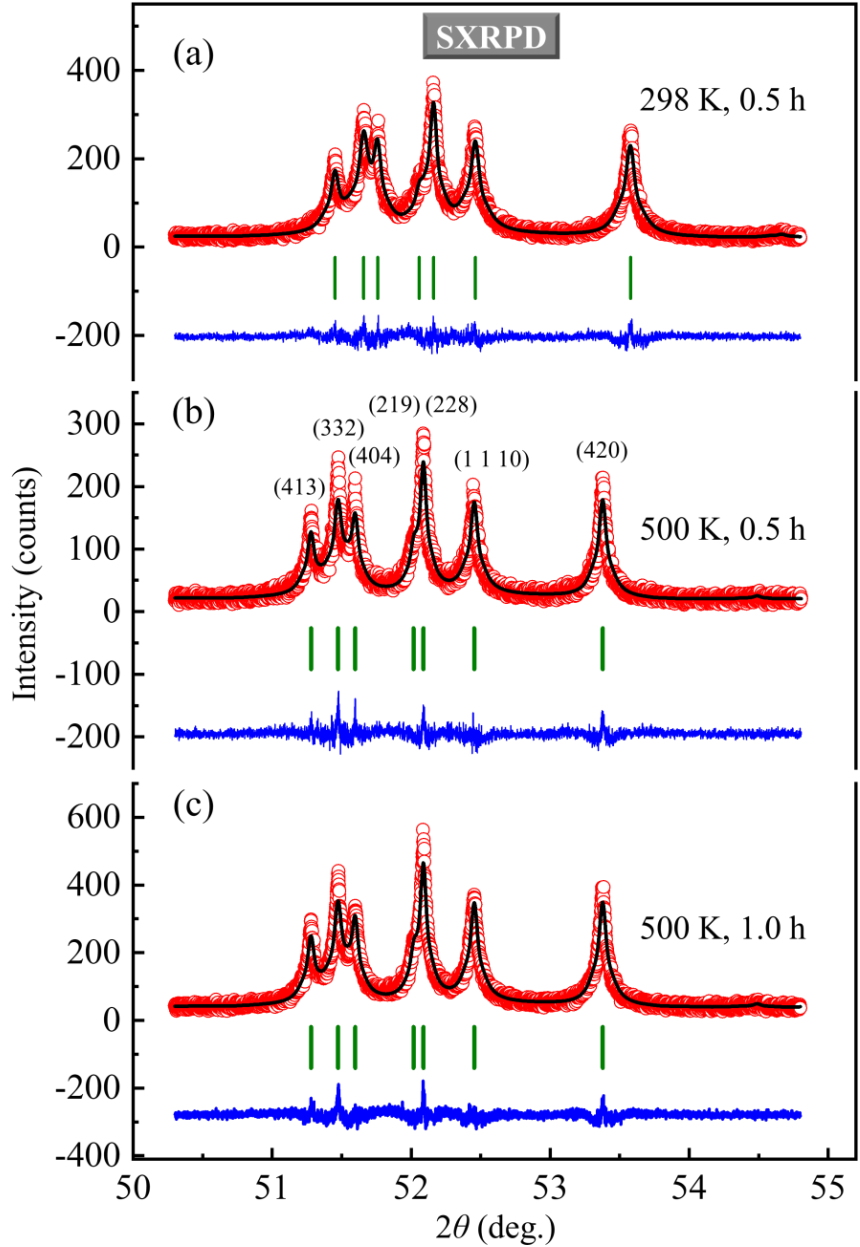

FIG. 4. For a clear comparison, we show observed (circles) and calculated (solid lines) synchrotron X-Ray powder diffraction patterns of the pulverized ErPd$_2$Si$_2$ single crystal in a small 2$\theta$ range of 50.3–54.8°. The patterns were collected at (a) 298 K, 0.5 h; (b) 500 K, 0.5 h; and (c) 500 K, 1.0 h. Vertical bars mark the positions of Bragg reflections. The bottom curves represent the difference between observed and calculated patterns. The Bragg peaks of (4 1 3), (3 3 2), (4 0 4), (2 1 9), (2 2 8), (1 1 10), and (4 2 0) were marked.

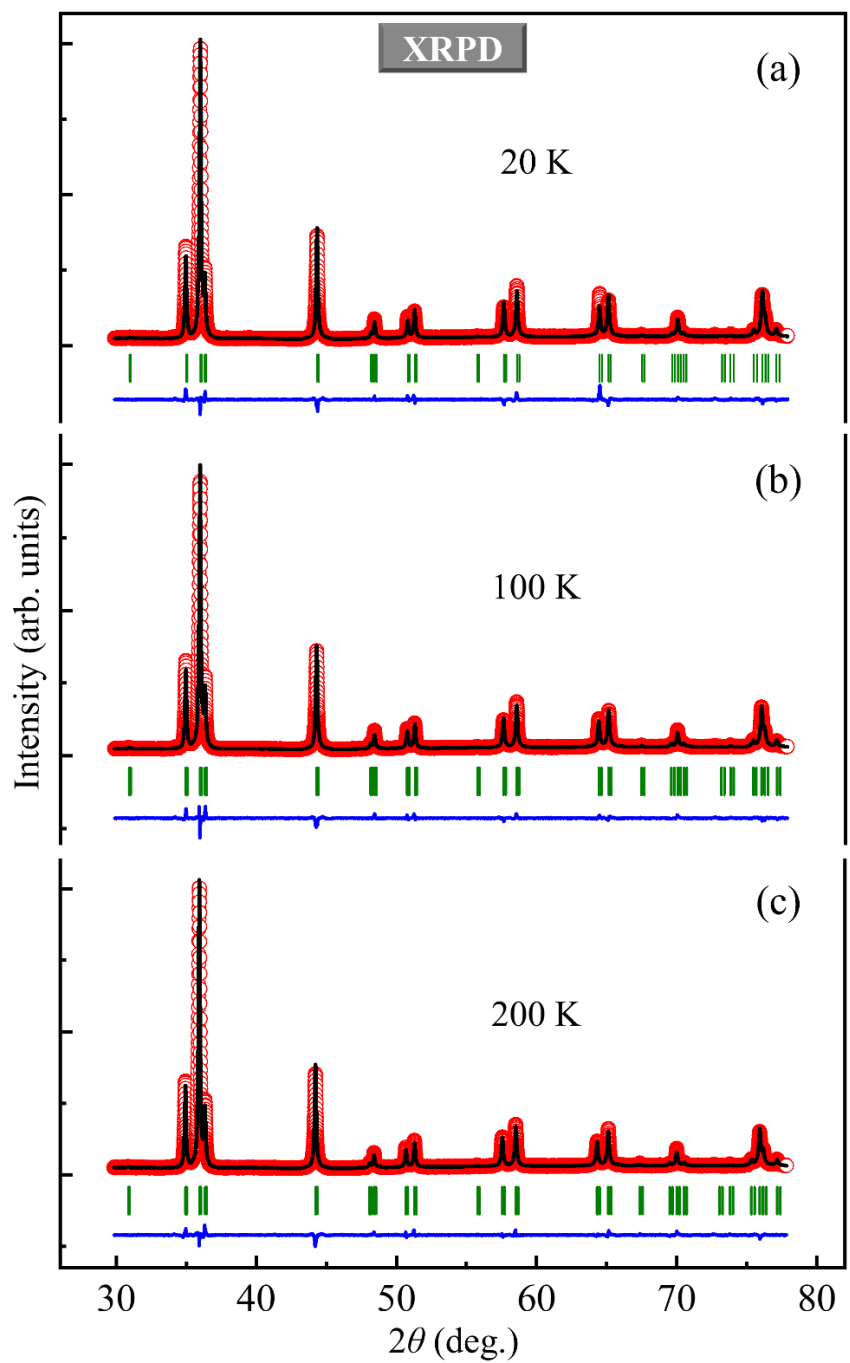

FIG. 5. Observed (circles) and calculated (solid lines) in-house X-Ray powder diffraction patterns of the pulverized ErPd$_2$Si$_2$ single crystal collected at 20 K (a), 100 K (b), and 200 K (c). Vertical bars mark the positions of Bragg reflections. The bottom curves represent the difference between observed and calculated patterns. The diffraction angle $2\theta$ range is from 30 to 78°.

TABLE I. Experimental conditions and refinement parameters for the SXRPD study of single-crystal ErPd$_2$Si$_2$ compound, including reliability values ($R_p$, $R_{wp}$, and $R_{exp}$) and goodness of fit ($\chi^2$). During refinements, we tried to include a general strain broadening model (quartic form), and the corresponding parameters were also listed.

| $T$ (K) | 298 | 500 | 500 |
| --- | --- | --- | --- |
| Counting time (h) | 0.5 | 0.5 | 1.0 |
| $R_p$ | 7.79 | 8.95 | 7.44 |
| $R_{wp}$ | 10.20 | 11.70 | 9.43 |
| $R_{exp}$ | 7.70 | 9.12 | 6.48 |
| $\chi^2$ | 1.74 | 1.64 | 2.12 |
| $R_p$ (strain) | 7.54 | 8.67 | 6.99 |
| $R_{wp}$ (strain) | 9.92 | 11.50 | 8.97 |
| $R_{exp}$ (strain) | 7.73 | 9.13 | 6.51 |
| $\chi^2$ (strain) | 1.65 | 1.59 | 1.90 |

TABLE II. Refined unit cell parameters of the intermetallic ErPd$_2$Si$_2$ single crystal from a SXRPD study, including lattice constants $a$ and $c$, unit-cell volume $V$, calculated density $D_{cal}$, atomic positions, and isotropic thermal parameter ($B$), obtained by FULLPROF refinements of SXRPD data collected on the I11 beamline (Diamond, UK) at 298 and 500 K. The Wyckoff sites of all atoms were listed. The numbers in parenthesis are the estimated standard deviations of the (next) last significant digit.

| SXRPD study of a pulverized ErPd$_2$Si$_2$ single crystal (Tetragonal, space group: *I4/mmm*) | | |
| --- | --- | --- |
| $T$ (K) | 298 | 500 |
| Counting time (h) | 0.5 | 0.5 |
| $a$ (Å) | 4.10320(2) | 4.11737(2) |
| $c$ (Å) | 9.88393(5) | 9.88143(5) |
| $V$ (Å$^3$) | 166.408(1) | 167.517(2) |
| $D_{cal}$ (g•cm$^{-3}$) | 8.706 | 8.649 |
| Er: | 2a: (0, 0, 0) | |
| $B$ (Er) (Å$^2$) | 0.823(11) | 1.183(7) |
| Pd: | 4d: (0, 0.5, 0.25) | |
| $B$ (Pd) (Å$^2$) | 0.860(11) | 1.256(9) |
| Si: | 4e: (0, 0, $z$) | |
| $z$ (Si) | 0.38064(14) | 0.37984(13) |
| $B$ (Si) (Å$^2$) | 1.200(27) | 1.226(25) |

TABLE III. Powder diffraction data of ErPd$_2$Si$_2$ from the SXRPD study with $\lambda$ = 0.827032 Å at 298 K.

| h | k | l | $2\theta_{obs}$ (°) | $d_{obs}$ (Å) | $I_{obs}$ | $I_{cal}$ |
|---|---|---|---|---|---|---|
| 0 | 0 | 2 | 9.600  | 4.942 | 2.03   | 2.12   |
| 1 | 0 | 1 | 12.529 | 3.790 | 29.51  | 27.27  |
| 1 | 1 | 0 | 16.388 | 2.901 | 0.10   | 0.08   |
| 1 | 0 | 3 | 18.526 | 2.569 | 30.59  | 29.18  |
| 1 | 1 | 2 | 19.026 | 2.502 | 100.00 | 100.00 |
| 0 | 0 | 4 | 19.267 | 2.471 | 15.94  | 16.26  |
| 2 | 0 | 0 | 23.256 | 2.052 | 34.80  | 35.33  |
| 2 | 0 | 2 | 25.211 | 1.895 | 0.44   | 0.42   |
| 1 | 1 | 4 | 25.396 | 1.881 | 2.73   | 2.71   |
| 2 | 1 | 1 | 26.500 | 1.804 | 8.53   | 8.95   |
| 1 | 0 | 5 | 26.853 | 1.781 | 11.57  | 11.56  |
| 0 | 0 | 6 | 29.076 | 1.647 | 0.11   | 0.11   |
| 2 | 1 | 3 | 29.896 | 1.603 | 15.09  | 15.57  |
| 2 | 0 | 4 | 30.375 | 1.578 | 19.34  | 18.00  |
| 2 | 2 | 0 | 33.123 | 1.451 | 11.97  | 12.29  |
| 1 | 1 | 6 | 33.556 | 1.433 | 17.07  | 16.87  |
| 2 | 2 | 2 | 34.564 | 1.392 | 0.11   | 0.11   |
| 3 | 0 | 1 | 35.543 | 1.355 | 1.70   | 1.81   |
| 2 | 1 | 5 | 35.814 | 1.345 | 10.58  | 9.70   |
| 1 | 0 | 7 | 36.084 | 1.335 | 1.96   | 2.05   |
| 3 | 1 | 0 | 37.168 | 1.298 | 0.07   | 0.06   |
| 2 | 0 | 6 | 37.559 | 1.284 | 0.20   | 0.18   |
| 3 | 0 | 3 | 38.217 | 1.263 | 3.40   | 3.47   |
| 3 | 1 | 2 | 38.476 | 1.255 | 23.56  | 23.22  |
| 2 | 2 | 4 | 38.603 | 1.251 | 8.57   | 8.38   |
| 0 | 0 | 8 | 39.108 | 1.235 | 3.29   | 3.34   |
| 3 | 1 | 4 | 42.195 | 1.149 | 0.88   | 0.84   |
| 1 | 1 | 8 | 42.665 | 1.137 | 0.03   | 0.03   |
| 3 | 2 | 1 | 42.909 | 1.131 | 1.91   | 2.00   |
| 3 | 0 | 5 | 43.141 | 1.125 | 2.65   | 2.54   |
| 2 | 1 | 7 | 43.373 | 1.119 | 2.35   | 2.28   |
| 2 | 2 | 6 | 44.645 | 1.089 | 0.10   | 0.08   |
| 3 | 2 | 3 | 45.217 | 1.076 | 3.91   | 3.97   |
| 1 | 0 | 9 | 45.883 | 1.061 | 0.72   | 0.68   |
| 2 | 0 | 8 | 45.996 | 1.058 | 7.86   | 7.72   |
| 4 | 0 | 0 | 47.546 | 1.026 | 3.32   | 3.41   |
| 3 | 1 | 6 | 47.868 | 1.019 | 10.25  | 9.98   |
| 4 | 0 | 2 | 48.624 | 1.004 | 0.03 | 0.03 |
| 4 | 1 | 1 | 49.369 | 0.990 | 1.21 | 1.23 |
| 3 | 2 | 5 | 49.578 | 0.986 | 3.25 | 3.12 |
| 3 | 0 | 7 | 49.786 | 0.982 | 0.70 | 0.69 |
| 3 | 3 | 0 | 50.627 | 0.967 | 0.02 | 0.02 |
| 4 | 1 | 3 | 51.452 | 0.953 | 2.47 | 2.44 |
| 3 | 3 | 2 | 51.657 | 0.949 | 4.07 | 4.09 |
| 4 | 0 | 4 | 51.758 | 0.947 | 2.98 | 2.92 |
| 2 | 1 | 9 | 52.057 | 0.942 | 0.92 | 0.92 |
| 2 | 2 | 8 | 52.160 | 0.941 | 5.04 | 4.94 |
| 1 | 1 | 10 | 52.461 | 0.936 | 4.04 | 4.07 |
| 4 | 2 | 0 | 53.577 | 0.918 | 4.43 | 4.44 |
| 4 | 2 | 2 | 54.568 | 0.902 | 0.05 | 0.03 |
| 3 | 3 | 4 | 54.665 | 0.901 | 0.15 | 0.15 |
| 3 | 1 | 8 | 55.052 | 0.895 | 0.03 | 0.03 |
| 4 | 1 | 5 | 55.447 | 0.889 | 2.06 | 2.00 |
| 3 | 2 | 7 | 55.639 | 0.886 | 0.98 | 0.97 |
| 1 | 0 | 11 | 56.213 | 0.878 | 0.72 | 0.74 |
| 4 | 0 | 6 | 56.704 | 0.871 | 0.03 | 0.03 |
| 4 | 2 | 4 | 57.471 | 0.860 | 4.09 | 4.09 |
| 3 | 0 | 9 | 57.749 | 0.856 | 0.33 | 0.32 |
| 3 | 3 | 6 | 59.446 | 0.834 | 2.22 | 2.26 |
| 0 | 0 | 12 | 60.271 | 0.824 | 0.43 | 0.42 |
| 4 | 3 | 1 | 60.746 | 0.818 | 0.59 | 0.60 |
| 5 | 0 | 1 | 60.746 | 0.818 | 0.29 | 0.29 |
| 4 | 1 | 7 | 61.109 | 0.813 | 0.72 | 0.68 |
| 2 | 1 | 11 | 61.650 | 0.807 | 1.08 | 1.07 |
| 4 | 2 | 6 | 62.114 | 0.802 | 0.07 | 0.05 |
| 4 | 3 | 3 | 62.569 | 0.796 | 1.19 | 1.13 |
| 5 | 0 | 3 | 62.569 | 0.796 | 0.57 | 0.55 |
| 5 | 1 | 2 | 62.751 | 0.794 | 3.92 | 3.84 |
| 1 | 1 | 12 | 62.917 | 0.792 | 0.07 | 0.06 |
| 3 | 2 | 9 | 63.104 | 0.790 | 0.47 | 0.47 |
| 4 | 0 | 8 | 63.195 | 0.789 | 2.37 | 2.29 |
| 3 | 1 | 10 | 63.461 | 0.786 | 4.17 | 3.97 |
| 5 | 1 | 4 | 65.427 | 0.765 | 0.13 | 0.13 |
| 2 | 0 | 12 | 65.503 | 0.764 | 1.28 | 1.23 |
| 5 | 2 | 1 | 65.957 | 0.760 | 0.34 | 0.34 |

*The intensity of Bragg (1 1 2) peak that has the strongest intensity in the whole pattern was normalized to 100, and those of other Bragg peaks were scaled accordingly.*

TABLE IV. Extracted unit cell parameters of the intermetallic ErPd$_2$Si$_2$ single crystal from a XRPD study, including lattice constants $a$ and $c$, unit-cell volume $V$, calculated density $D_{cal}$, atomic positions, and isotropic thermal parameter ($B$), obtained by FULLPROF refinements of XRPD data collected on an in-house X-ray diffractometer at 20, 100, and 200 K. The numbers in parenthesis are the estimated standard deviations of the (next) last significant digit.

| | XRPD study of a pulverized ErPd$_2$Si$_2$ single crystal (Tetragonal, space group: $I4/mmm$) | | |
|---|---|---|---|
| $T$ (K) | 20 | 100 | 200 |
| $a$ (Å) | 4.08607(4) | 4.08878(4) | 4.09316(4) |
| $c$ (Å) | 9.89253(12) | 9.88694(10) | 9.88116(10) |
| $V$ (Å$^3$) | 165.165(3) | 165.291(3) | 165.548(3) |
| $D_{cal}$ (g·cm$^{-3}$) | 8.772 | 8.765 | 8.752 |
| Er: | | 2a: (0, 0, 0) | |
| $B$ (Er) (Å$^2$) | 2.01(4) | 2.38(3) | 2.35(3) |
| Pd: | | 4d: (0, 0.5, 0.25) | |
| $B$ (Pd) (Å$^2$) | 2.39(4) | 2.75(3) | 3.22(3) |
| Si: | | 4e: (0, 0, $z$) | |
| $z$ (Si) | 0.37655(24) | 0.37631(23) | 0.37653(23) |
| $B$ (Si) (Å$^2$) | 2.49(7) | 3.97(7) | 4.25(7) |
| $R_p$ | 3.35 | 3.08 | 2.96 |
| $R_{wp}$ | 4.99 | 4.20 | 4.11 |
| $R_{exp}$ | 3.45 | 3.51 | 3.49 |
| $\chi^2$ | 2.09 | 1.43 | 1.38 |